\documentstyle[aps,12pt,epsfig]{revtex}

\parskip=\medskipamount

\def\be{\begin{equation}}
\def\ee{\end{equation}}
\def\disp{\displaystyle}

\def\lam{\lambda}
\def\R{{\sf I\kern-.15em R}}
\def\T{{\sf T\kern-.45em T}}
\def\C{\kern.1em{\raise.47ex\hbox{$\scriptscriptstyle |$}}
             \kern-.40em{\sf C}}
\def\Z{{\sf Z\kern-.45em Z}}

\def\al{\alpha}
\def\slz{PSL(2,\Z)}
\def\slr{PSL(2,\R)}
\def\hfl#1#2{\smash{\mathop{\hbox to 
10mm{\rightarrowfill}}\limits^{\scriptstyle#1}_{\scriptstyle#2}}}

\begin{document}


\title
{Random walks on hyperbolic groups and their Riemann surfaces}

\author{Sergei Nechaev$^{1,2}$ and Rapha\"el
Voituriez$^{1}$}

\address{{\it $^{1}$Laboratoire de Physique Th\'eorique et Mod\`eles
Statistiques, Universit\'e Paris Sud, \\ 91405 Orsay Cedex, France\\$^{2}$ L D
Landau Institute for Theoretical Physics, 117940, Moscow, Russia}}

\maketitle


\begin{abstract}

We investigate invariants for random elements of different hyperbolic groups.
We provide a method, using Cayley graphs of groups, to compute the probability
distribution of the minimal length of a random word, and explicitly compute the
drift in different cases, including the braid group $B_3$.  We also compute in
this case the return probability.  The action of these groups on the hyperbolic
plane is investigated, and the distribution of a geometric invariant, the
hyperbolic distance, is given.  These two invariants are shown to be related by
a closed formula.

\end{abstract}

\bigskip

\section{Introduction}
\label{sect1}

This paper is devoted to a systematic study of random walks on the
modular group $\slz$ and on some closely related groups: the braid group
$B_3$, the Hecke groups $H_q$ and the free groups $F_n$ (all definitions
are given below). We study simultaneously the limiting distribution of random walks on the Cayley graphs of these
groups and on their Riemann surfaces. We analyze the statistical
properties of random walks on the Cayley graphs of the
above mentioned groups both in  a metric of words and in  the  natural metric of the hyperbolic plane.

The very subject of our investigation is not new---the statistics of
Markov chains on the subgroups of the group $\slr$ has been extensively
studied in the mathematical literature.  Among the known results
connected to the theme of our work we can mention: (a) the central limit
theorem for Markov multiplicative processes on discrete subgroups of the
group $\slr$ \cite{terras,doob}, (b) some particular examples of exact results for limiting
distribution functions of random walks on Cayley graphs of free and
modular groups \cite{voitnech,woess,woess2,ne_sem_kol} and (c) conjectures concerning the return probability and
drift on the braid group $B_3$  \cite{vershik,nech_gros}.

In the present work we rigorously compute the drift and the return
probability for symmetric random walks (in metric of words) on the
groups $H_q$ and $B_3$.  Moreover, as it has been said, we pay a special attention to the
statistics of random walks on the Riemann surfaces of the groups
$\slz,B_3,H_q,F_n$.  Namely, we study a $2\times2$   matrix representation of these groups and consider their homographic  action\footnote{since the 2--representation of $B_3$ is not unimodular, this action is not faithfull} on the hyperbolic plane ${\cal H}$.  This allows to embed the Cayley graphs in ${\cal H}$,  and to define isometric hyperbolic lattices. Taking advantage of the hyperbolic metric on ${\cal H}$, we investigate the probability distribution of the 
geodesic distance between ends of random processes with symmetric transition probabilities on these lattices of ${\cal H}$.  We show that this problem reduces to the study of the modulus of a random product of matrices. The part of our
investigation is semi--analytic and is based on numerical results on the
structure of the invariant distribution of geodesics at the boundary of
${\cal H}$.  We found very interesting the fact that the drift on a
Cayley graph in  a metric of words coincides after proper normalization with the drift on the corresponding isometric lattice of ${\cal H}$ in the  natural hyperbolic metric. This result establishes a nontrivial relation between two group invariants: in one hand the irreducible length of an element, which does not depend on the representation, and on the other hand, the hyperbolic distance associated to an element (directly linked to its modulus), defined only for this matrix representation.

As an application of our results,  we consider the relation between the
distribution of Alexander knot invariants and the asymptotic behavior of
random walks over the elements of the simplest nontrivial braid group
$B_3$. This class of problems arises naturally even beyond the aims of
our particular investigation: the limiting behavior of Markov chains on
braid and so-called "local" groups can be regarded as a first step in a
consistent development of harmonic analysis on branched manifolds
(Teichm\"uller spaces are an example).

The paper is structured as follows. In section \ref{sec:2} we give the basic definitions and introduce the different groups and their Cayley graphs. A general solution of the diffusion problem on these graphs, as well as exact computations of the drift and the return probability for $B_3$ are developed in section \ref{sec:3}. Section \ref{sec:4} is devoted to the study of the action of these groups in the hyperbolic plane; a discussion of our results and the relation between the different approaches are presented in section \ref{sec:5}.

\section{Hyperbolic groups $\slz,\,H_q,\, B_3,\, F_n$ and their 
Cayley graphs}
\label{sec:2}

\subsection{Basic definitions}

We consider a special class of so-called hyperbolic groups -- the modular group $\slz$, and some of its
generalizations -- the Hecke groups $H_q$ and the braid group $B_3$.  We also
recall already known properties of the free groups $F_n$, usefull in the
context of our work.

1. The modular group $\slz$ is a free product ${\Z}_2\star{\Z}_3$ of two
cyclic groups of 2nd (generated by $a_2$) and 3rd (generated by $b_3$ )
orders. In a standard framing using  generators $S$ (inversion) and
$T$ (translation), the group $\slz$ is defined by the following relations
\be \label{eq:1}
\begin{array}{r}
(ST)^3=b_{3}^3=1 \\ S^2=a_{2}^2=1
\end{array}
\ee
Being a discrete subgroup of the group $\slr$, the generators $T$ and $S$ of 
the modular 
group $\slz$ have a natural representation by unimodular matrices $\hat{T}$ 
and $\hat{S}$:
\be \label{eq:2}
\hat{T}=\left(\begin{array}{cc} 1 & 1\\ 0 & 1
\end{array}\right); \qquad
\hat{S}=\left(\begin{array}{cc} 0 & 1 \\ -1 & 0
\end{array}\right)
\ee

2. In addition to the modular group $\slz$ we shall consider the
so-called Hecke group $H_q$ which ``interpolates'' between the modular
group (for $q=3$) and the free group $F_3$ with 3 generators, the
so-called $\Lambda$ group (for $q=\infty$). The Hecke group $H_q$ is
isomorphic to ${\Z}_2\star{\Z}_q$ (we denote by $a_2$ and $b_q$ the generators of orders 2 and $q$). It  is defined by
straightforward generalization of the relations (\ref{eq:1})
\be \label{eq:1a}
\begin{array}{r}
(ST_q)^q=b_{q}^q=1 \\ S^2=a_{2}^2=1
\end{array}
\ee
and the generators $T_q$ and $S$ have the following matrix representation
(compare to (\ref{eq:2})):
\be \label{eq:2a}
\hat{T}_q=\left(\begin{array}{cc} 1 & 2\cos\frac{\pi}{q} \\ 0 & 1
\end{array}\right); \qquad
\hat{S}=\left(\begin{array}{cc} 0 & 1 \\ -1 & 0
\end{array}\right)
\ee
The parameter $q$ takes the discrete values $q=3,4,5,6,\ldots$. 

\begin{figure}[ht]
\begin{center}
\epsfig{file=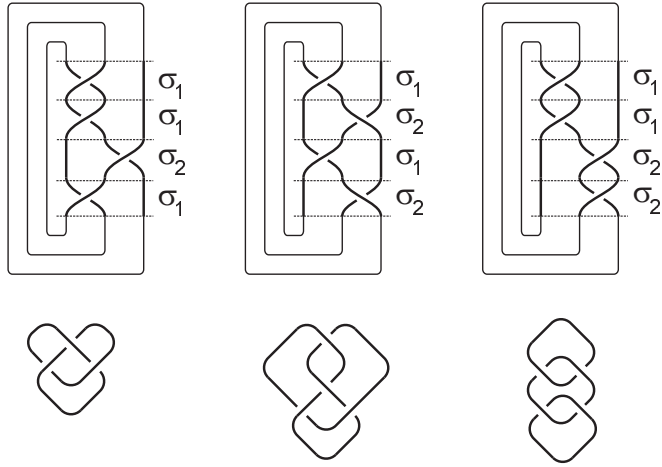,width=10cm}
\end{center}
\caption{Some examples of closed braids and corresponding links.}
\label{fig:1}
\end{figure}

3. The braid group $B_3$ is defined by the following commutation relations
among generators $\{\sigma_1,\sigma_2\}$:
\be \label{eq:3}
\begin{array}{l}
\sigma_1\sigma_2\sigma_1=\sigma_2\sigma_1\sigma_2 \\
\sigma_1\sigma_1^{-1}=\sigma_2\sigma_2^{-1}=e
\end{array}
\ee
In our further construction we shall repeatedly use the following framing:
\be
\begin{array}{l}
{\tilde a}=\sigma_1\sigma_2\sigma_1\\
{\tilde b}=\sigma_1^{-1}\sigma_2^{-1}
\end{array}
\ee
The generators of the group $B_3$ can be represented by 
$PGL(2,{\R})$--matrices. To be more specific, the generators $\sigma_1$
and $\sigma_2$ in the Magnus representation \cite{birman} read 
\be \label{eq:4}
\hat{\sigma}_1=\left(\begin{array}{cc} -t & 1 \\ 0 & 1
\end{array}\right); \qquad
\hat{\sigma}_2=\left(\begin{array}{cc} 1 & 0 \\ t & -t
\end{array}\right)
\ee
where $t$ is the free parameter. We conveniently introduce the parameter 
$u=\sqrt{-t}$, and consider normalized generators of determinant 1:

\be \label{eq:4bis}
{}_{u}\hat{\sigma}_1=\left(\begin{array}{cc} u & 1/u \\ 0 & 1/u
\end{array}\right); \qquad
{}_{u}\hat{\sigma}_2=\left(\begin{array}{cc} 1/u & 0 \\ -u & u
\end{array}\right)
\ee
The  group generated by ${}_{u}\hat{\sigma}_1$ and ${}_{u}\hat{\sigma}_2$ will be denoted later on as $\slz_u$. Indeed it is just a deformation of $\slz$, which preserves all its commutation relations. For $u=1$, one has $\slz_u=\slz$, and  the group $B_3$ is a
central  extension of $\slz_u$ of center
\be \label{eq:5}
(\hat{\sigma}_1\hat{\sigma}_2)^{3\lambda}=(\hat{\sigma}_2\hat{\sigma}_1)^{3\lambda}=
(\hat{\sigma}_1\hat{\sigma}_2\hat{\sigma}_1)^{2\lambda}=
(\hat{\sigma}_2\hat{\sigma}_1\hat{\sigma}_2)^{2\lambda}=
\left(\begin{array}{cc}
t^{3\lambda} & 0 \\ 0 & t^{3\lambda}
\end{array}\right),\  \forall\lambda\in\Z
\ee
(let us note that the center is isomorphic to $\Z$).
Recall that graphically, to each word of $B_3$ correspond a particular three--strand braid, going from
above downwards. (see Fig.\ref{fig:1}). A closed braid is obtained by gluing the "top" and "bottom" free ends on a cylinder. Any closed braid defines a link (in particular,
a knot). However the correspondence between braids and knots (links) is
not one--to--one and each link (knot) can be represented by infinite
number of different braids (see \cite{kauf,birman}). The irreducible length of a braid gives nevertheless an interesting characteristic of the link complexity.

There exists an extensive literature on general properties of braid
groups---see \cite{birman}; for the last works on the normal forms of words, we
shall quote \cite{birman2}. 

Any element of the group $G=\{\slz,H_q,B_3\}$ is defined by a word in 
alphabets of corresponding letters (generators): 
\begin{itemize}
\item $\{S,T,T^{-1}\}$ or $\{a_2,b_3,b_{3}^{-1}\}$---for
$\slz$
\item $\{S,T_q,T_q^{-1}\}$ or $\{a_2,b_q,b_{q}^{-1}\}$---for $H_q$ 
\item $\{\sigma_1,\sigma_2, \sigma_1^{-1},\sigma_2^{-1}\}$ or 
$\{\tilde{a},\tilde{a}^{-1},\tilde{b},\tilde{b}^{-1}\}$---for $B_3$. 
\end{itemize}
We denote by $w_n$ a word corresponding to a given record of length $n$,
and by $L^{G}(w_n)$ the irreducible length in the metric of words (the superscript $G$ is precised only when it is necessary), or in other terms the minimal number of generators necessary to build $w_n$. The irreducible length can be also viewed as a distance from the unity on the Cayley graph of the group $G$.
Note that $L^{G}(w)$ depends on the set of generators we consider.

\subsection{Cayley graphs}

\begin{figure}[ht]
\begin{center}
\epsfig{file=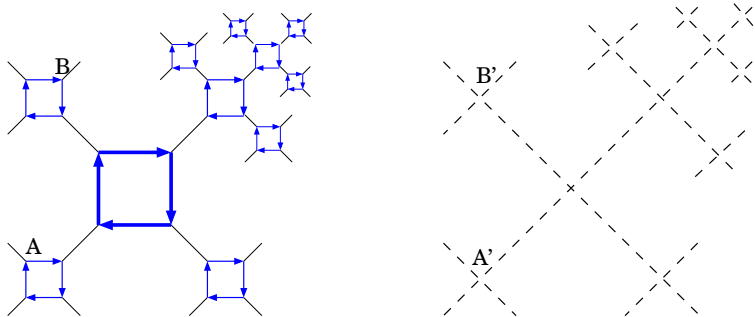,width=10cm}
\end{center}
\caption{Cayley graph of $H_q$, with $q=4$. Arrows correspond to
generator $b_q=ST_q$, thin lines to generator $a_2=S=S^{-1}$. The dashed
graph is the corresponding backbone graph, a tree $\T_q$, graph of $F_q$.
The distance $d$ between A and B is $d=5$, while the distance $k$
(corresponding to the distance between A' and B') is $k=2$.}
\label{fig:2}
\end{figure}

The modular group $\slz$ is a particular case of the Hecke group $H_q$ at
$q=3$. We therefore consider without any loss of generality the Cayley graphs of the groups $H_q$ for $q=3,4,...$. The Cayley graph of $B_3$ will be constructed afterwards. We investigate in this part only the abstract presentation of the groups in terms of commutation relations and do not pay attention to any representation. We recall that the Cayley graph of a group $G$ is the graph whose vertices
are labeled by group elements, and whose links are as follows: $w$ and
$w'$ are linked if and only if there exists a generator $g$ such that
$w'=wg$. Following this rule, we can easily construct the Cayley graph ${\cal G}_q$ of the group $H_q$ represented by $\{a_2,b_q,b_{q}^{-1}\}$. For any finite
values of $q$ the graph ${\cal G}_q$ has local $q$--cycles (because $b_q$ is of order $q$), while the corresponding dual (or "backbone") graph is the tree graph $\T_q$, which is precisely the graph of $F_q$. This is due to the free product structure of $H_q\sim\Z_2\star\Z_q$ (see explanations below). The graph ${\cal G}_q$ is shown in Fig.\ref{fig:2}, where the backbone graph is marked by a dotted line.


\section{Diffusion on graphs}
\label{sec:3}
In this section we investigate some statistical properties of random walks on 
the groups introduced above, using their Cayley graphs. In particular  we consider {\it simple random walks}, that are walks of nearest neighbour  type with symmetric transition probabilities.

\subsection{Random walk on $PSL(2,\Z)$ and $ H_q$}\label{sec3a}

We consider free product groups of the form $\Z_2\star\Z_q$ (isomorphic
to $H_q$), in the framing which uses the generators $a_2$ and $b_q$, of
order $2$ and $q$ respectively.  Graphs of such groups are shown in Fig.\ref{fig:2}. We define two different ``metrics'' of words on those graphs. The first
metric is associated with the geodesic distance $d$ on the graph---the
minimal number of steps between two points, and the second metric is
associated with the geodesic distance $k$ (called later the "generation")
on the backbone graph $\T_q$. Our goal is
to compute the probability $P_q(d,n)$ of being at a distance $d$ from the
initial point (the root of the graph) after $n$ random steps. The
probability ${\bar P}_q(k,n)$ of being on the  backbone graph at a
generation $k$ from a root point after $n$ random steps will also be of
use.

First of all we compute $P_3(d,n)$ and ${\bar P}_3(k,n)$ for the case of 
$PSL(2,\Z)$. In this case the graph structure ensures the relation 
$$
P_3(2x,n)={\bar P}_3(x,n)
$$ 
Therefore we can consider only ${\bar P}_3(k,n)$.  Write 
$$
{\bar P}_3(k,n)={\bar P}_{3}^{i}(k,n)+{\bar P}_{3}^{o}(k,n)
$$
distinguishing for  an elementary  triangular cell located at  generation $k$ the vertex closest to the root (corresponding to  ${\bar
P}_{3}^{i}(k,n)$) and the two others (corresponding to ${\bar
P}_{3}^{o}(k,n)$) (see Fig.\ref{fig:7}).  A direct enumeration gives the 
following master equation for $k\ge2$:
\be\label{maeq}
\left\{\begin{array}{l}
{\bar P}_{3}^{i}(k,n+1)=\frac{1}{3}\left({\bar P}_{3}^{o}(k,n)+{\bar
P}_{3}^{o}(k-1,n)\right)\\ {\bar P}_{3}^{o}(k,n+1)=\frac{2}{3}{\bar
P}_{3}^{i}(k,n)+\frac{1}{3}\left({\bar P}_{3}^{o}(k,n)+{\bar
P}_{3}^{i}(k+1,n)\right)
\end{array}\right.
\end{equation}
with initial conditions of the form
\be\left\{\begin{array}{l}
{\bar P}_{3}^{i}(k,0)=\alpha\delta_{k,0}\\
{\bar P}_{3}^{o}(k,0)=(1-\alpha)\delta_{k,0}
\end{array}\right.
\end{equation}
where $\alpha$ is an arbitrary parameter fixing the initial condition and
varying in the interval $0\le \alpha \le 1$. We are seeking for the
asymptotic ($1\ll k\le n$) solution to (\ref{maeq}) near the
maximum of the probability distribution, and therefore will not take into
account the specific form of the boundary condition.

Define the Laplace--Fourier transform:
\be
Q^{i,o}(x,s)={\cal T}\left[{\bar P}_{3}^{i,o}\right]=\sum_{n=0}^{\infty}s^n
\sum_{k=-\infty}^{+\infty}e^{ikx}{\bar P}_{3}^{i,o}(k,n)
\ee
whose inverse can be written in the form
\be\label{inv}
{\bar P}_{3}^{i,o}(k,n)=\frac{1}{4i\pi^2}\oint\frac{ds}{s^{n+1}}
\int_{-\pi}^{\pi}e^{-ikx}Q^{i,o}(x,s)dx
\ee
One straightforwardly obtains the following algebraic system of linear equations:
\be\label{alsys}
\left\{\begin{array}{l}
\disp Q^{i}(x,s)-\frac{s}{3}(1+e^{ix})Q^{o}(x,s)=\alpha\\
\disp -\frac{s}{3}(2+e^{-ix})Q^{i}(x,s)+(1-\frac{s}{3})Q^{o}(x,s)=1-\alpha
\end{array}\right.
\end{equation}
which determines the function $Q^{i,o}(x,s)$:
\be
\disp Q^{i,o}(x,s)=\frac{a_{\alpha}^{i,o}(x)+b_{\alpha}^{i,o}(x)s}{\disp
s^2+\frac{3}{3+e^{-ix}+2e^{ix}}s-\frac{9}{3+e^{-ix}+2e^{ix}}}=
\frac{a_{\alpha}^{i,o}(x)+b_{\alpha}^{i,o}(x)s}{p(x,s)}
\ee
where
\be\left\{\begin{array}{l}
\disp a_{\alpha}^{i}(x)=\frac{9\al}{3+e^{-ix}+2e^{ix}} \medskip \\
\disp a_{\alpha}^{o}(x)=\frac{9(1-\al)}{3+e^{-ix}+2e^{ix}} \medskip \\
\disp b_{\alpha}^{i}(x)=\frac{3(1-2\al+(1-\al)e^{ix})}{3+e^{-ix}+2e^{ix}} \medskip 
\\
\disp b_{\alpha}^{o}(x)=\frac{3\al(2+e^{-ix})}{3+e^{-ix}+2e^{ix}}
\end{array}\right.
\ee
Denote by $s_{\pm}(x)$ the roots of $p(x,s)$. Using (\ref{inv}) one can
rewrite
\be\label{iform}
{\bar P}_{3}^{i,o}(k,n)=
\frac{1}{2\pi}\int_{-\pi}^{\pi}dx\frac{e^{-ikx}}{s_{+}-s_{-}}
\left[a_{\alpha}^{i,o}(x)\left(\frac{1}{s_{+}^{n+1}}-\frac{1}{s_{-}^{n+1}}
\right)+b_{\alpha}^{i,o}(x)\left(\frac{1}{s_{+}^{n}}-\frac{1}{s_{-}^{n}}
\right)\right]
\ee
We are interested in the $n,k\gg 1$ regime, and therefore consider the
integrand in (\ref{iform}) for $x\to0$. Here we expose the second order
computation, keeping in mind that any order can be reached the same way.
With
\be
\disp\left\{\begin{array}{l}
\disp s_{+}=-\frac{3}{2}+\frac{3ix}{20}-\frac{51x^2}{250}+O(x^3) \medskip \\
\disp s_{-}=1-\frac{ix}{15}+\frac{209x^2}{2250}+O(x^3)
\end{array}\right.
\end{equation}
one gets
\be \label{ga}
\begin{array}{lll}
\disp {\bar P}_{3}(k,n)=\frac{1}{2\pi}\int_{-\pi}^{\pi}dx\;
e^{-ikx}(1-\frac{ix}{15}+\frac{209x^2}{2250})^{-n}&\approx & 
\disp \frac{1}{2\pi}\int_{-\infty}^{\infty}dx\;
\exp\left[-ikx-n\left(\frac{107x^2}{1125}-\frac{ix}{15}
\right)\right] \medskip \\ & \approx & 
\disp \frac{A}{n^{1/2}}\exp\left[-\frac{1125(k-\frac{n}{15})^2}{428n}\right]
\end{array}
\ee
where $A$ is the normalization constant.

The expression (\ref{ga}) allows one to compute the limiting value of the 
normalized drift $\bar{l}_3$
$$
\bar{l}_3=\lim_{n\to\infty} \frac{\left<k\right>_3}{n}
$$ 
on the backbone graph $\T_3$, where
$$
\disp\left<k\right>_3=\int_{-\infty}^{\infty}k \bar{P}_3(k,n) d 
k=\frac{n}{15}
$$
and, hence, the drift $l_3$
$$
l_3=\lim_{n\to\infty}\frac{\left<d\right>_3}{n}=2\bar{l}_3=\frac{2}{15}
$$
on the graph $H_3$.

Let us generalize these computations to the case of $H_q$. One can write
\be\label{pq}
\disp {\bar P}_q(k,n)=\sum_{i=1}^{[\frac{q}{2}]+1}{\bar P}_{q}^{i}(k,n)
\ee
and define the constants $\rho_i,\ 1\le i\le[\frac{q}{2}]+1$ (assuming the existence of corresponding 
limits): 
\be
\rho_{i}=\lim_{k\to\infty}\left(\lim_{n\to\infty}\frac{{\bar 
P}_{q}^{i}(k,n)}{{\bar P}_{q}(k,n)}\right)
\ee
which satisfy the normalization condition
\be
\sum_{i=1}^{[\frac{q}{2}]+1}\rho_{i}=1
\ee

\begin{figure}[ht]
\begin{center}
\epsfig{file=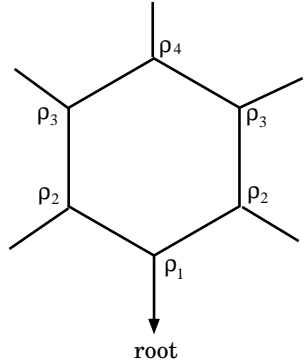,width=4cm}
\end{center}
\caption{Different types of vertices and their corresponding weights (here
$q=6$). $\rho_i$ gives the asymptotic ($k,n\to\infty$) probability of being at 
a vertex of type $i$.}
\label{fig:3}
\end{figure}

The sum (\ref{pq}) runs over $[\frac{q}{2}]+1$ non-equivalent
vertices (the graph is locally $\Z_2$ symmetric, see Fig.\ref{fig:3}) of
the elementary $q$--gone of the graph. Proceeding in the standard way,
we define the transform
\be
Q^{i}_q(x,s)={\cal T}\left[{\bar P}_{q}^{i}\right]
\ee
and derive the master equation, whose solution can be expressed in the 
following form
\be \label{eq:aa}
{\bar P}_{q}^{i}={\cal T}^{-1}\left[\sum_{j=1}^{n_q}
\alpha_{j}(M_{q}^{-1})^{ji}(x,s,\vec{\rho})\right]
\ee
where the $\alpha_j$ parametrize the initial conditions
\be \label{eq:ab}
\sum_{i=1}^{[\frac{q}{2}]+1}Q_{q}^{i}M_{q}^{ij}(x,s,\vec{\rho})=\alpha_j
\ee
and
\be \label{eq:ac}
M_{q}(x,s,\vec{\rho})=\disp \pmatrix{-1& \frac{s}{3}(1+e^{ix})&
\frac{s}{3}e^{ix} & \cdots & \cdots &\frac{s}{3}e^{ix} \cr
\frac{s}{3}(2+\frac{\rho_{2}}{1-\rho_1}e^{-ix})& -1 & \frac{s}{3}& 0 & \cdots &
0\cr \frac{s}{3}\frac{\rho_{3}}{1-\rho_1}e^{-ix}&\frac{s}{3}& -1 & 
\frac{s}{3} & \ddots & \vdots \cr
\frac{s}{3}\frac{\rho_{4}}{1-\rho_1}e^{-ix} & 0& \ddots &\ddots &\ddots & 0 \cr
\vdots&\vdots&\ddots&\frac{s}{3} & -1 &\frac{s}{3} \cr
\frac{s}{3}\frac{\rho_{n}}{1-\rho_1}e^{-ix} & 0 & \cdots & 0 &
\frac{s}{3}&-1+\frac{s}{3}\cr}
\ee

For $n,k\gg 1$ one obtains, using the same method as for $q=3$
\be
{\bar P}_{q}^{i}(k,n)=
{\bar \rho}_{i}(\vec{\rho})\delta(k-{\bar l}_q(\vec{\rho})n)
\ee
where
\be
{\bar l}_q(\vec{\rho})=\lim_{n\to\infty}\frac{\langle k\rangle_q}{n}=
i\frac{ds^{q}_{-}}{dx}
\ee
and $s^{q}_{-}$ is the  root of the polynomial ${\rm
det}\left(M_{q}(x,s,\vec{\rho})\right)$ the closest to zero.

To make the system of equations (\ref{eq:aa})--(\ref{eq:ac})
self--consistent we must set
\be
\rho_i={\bar \rho}_{i}(\vec{\rho})
\ee
which closes a system of equations determining $\rho_i$. Finally, one can 
write the limiting drift in the following form
\be
l_q=\lim_{n\to\infty} \frac{\left<d\right>_q}{n}=
{\bar l}_q(\vec{\rho})\left(1+\frac{\sum_{i=2}^{n_q}(i-1)\rho_i}{1-\rho_1}\right)
\ee
One can check that this formalism gives for $q=3$ the same results as has
been derived above.

\subsection{Random walk on $B_3$: drift and return probability}
\subsubsection{Analytic results}

We now focus on the braid group $B_3$ and in particular explain why
some statistical characteristics of random processes on $B_3$ have
the same asymptotic behavior as the ones on $PSL(2,\Z)$. The key point 
is that $B_3$  is a central extension of 
$PSL(2,\Z)$. Let us recall that the center $Z$ of $B_3$, generated by ${\tilde
a}^2={\tilde b}^3=\Delta^{2}$, is isomorphic to $\Z$. We denote by $\pi$ the
canonical quotient map
\be
\pi:\ B_3\longrightarrow \frac{B_3}{Z}\sim PSL(2,\Z)
\ee
One has then
\be\left\{\begin{array}{l}
\pi(\sigma_1)={\bar \sigma}_1=a_2b_3\\
\pi(\sigma_2)={\bar \sigma}_2=b_3a_2\\
\pi({\tilde a})=a_2\\
\pi({\tilde b})=b_3
\end{array}\right.
\ee
where $a_2$ and $b_3$ are defined in (\ref{eq:1})

\begin{figure}[ht]
\begin{center}
\epsfig{file=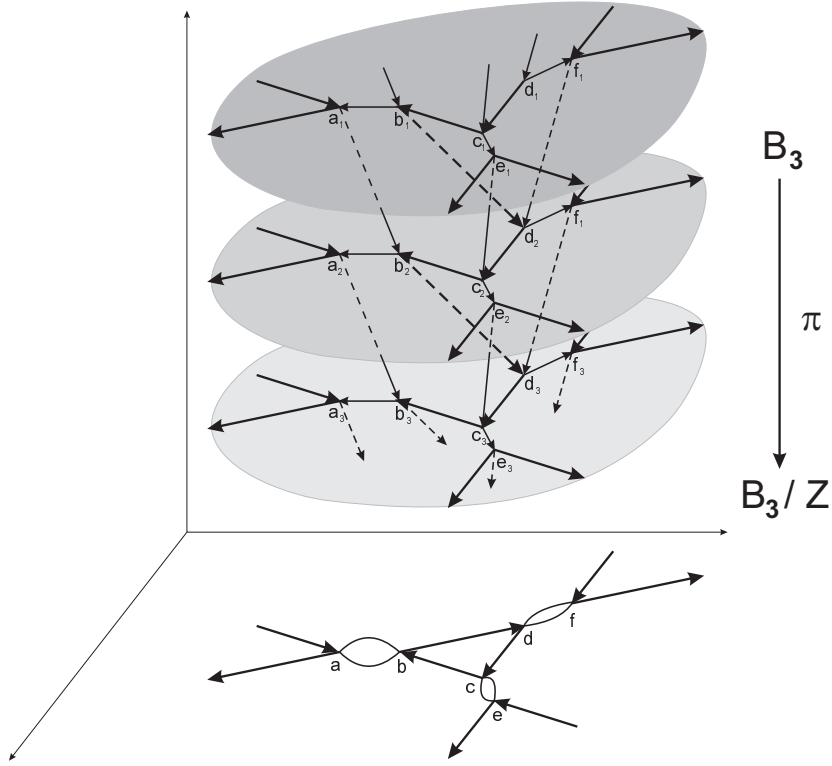,width=8cm}
\end{center}
\bigskip

\caption{$B_3$ Cayley graph and its projection ($\slz$). Thin arrows correspond
to 
${\tilde a}$, thick ones to ${\tilde b}$. Note that $\pi(\alpha_i)=\alpha$,  
$\pi(\beta_i)=\beta$ and so on. Recall that $a_2$ has to be identified with 
$a^{-1}_2$}
\label{fig:4}
\end{figure}

A natural representation of the Cayley graph of $B_3$ is three dimensional.  As
shown in Fig.\ref{fig:4}, the map $\pi$ can then be viewed as a projection from 3D to 2D.

Consider now an $n$--letter random word $w_n$ written in terms of generators of
the group $B_3$:
$$
\disp w_n=\prod_{i=1}^{n}\sigma_{r_i}
$$
where we set $\sigma^{-1}_{r_i}=\sigma_{-r_i}$ and indices $r_i$ are
uniformly distributed in $\{-2,-1,1,2\}$. We recall that  $L^{B_3}(w_n)$ is the irreducible  length of $w_n$. It is evident that 
\be\label{ineq1}
L^{B_3}(w_n)\ge L^{\slz}(\pi(w_n)) 
\ee
(keeping in mind the geometrical interpretation of $B_3$ shown in Fig.\ref{fig:4}, we can easily derive eq.(\ref{ineq1}) following from a 
triangular inequality). Consider now the irreducible decomposition
in $\slz$:
\be\label{dec}
\disp \pi(w_n)=\prod_{i=1}^{L^{\slz}(\pi(w_n))}{\bar \sigma}_{r'_i}
\ee
The asymptotic value of $\left<L^{\slz}(\pi(w_n))\right>$ for $n\gg 1$ is computed in 
appendix A by a straightforward adaptation of the method introduced in a 
previous section. From (\ref{dec}) and the definition of
the quotient map, we get:
\be\label{dec2}
\disp w_n={\Delta}^{2f(n)}\prod_{i=1}^{L^{\slz}(\pi(w_n))}\sigma_{r'_i}
\ee
where $\Delta^2$ is  6--letter word in the alphabet $\{\sigma_1,...,\sigma_2^{-1}\}$ (see Eq.\ref{eq:5}), what implies the following condition on $f(n)$:
$$
|f(n)|\le \frac{n}{6}
$$ 
Hence, the irreducible length of the word $w_n$ can be estimated from above
\be\label{ineq2}
L^{B_3}(w_n)\le 6|f(n)|+L^{\slz}(\pi(w_n))  
\ee
Let us show now that the Markovian process $f(n)$ is such that   
\be\label{on}
\left<|f(n)|\right>=O(\sqrt{n})
\ee

1. The symmetry of the process implies that the words $\Delta^{2}$ and 
${\Delta}^{-2}$ appear with same  probability, which gives $\left<f(n)\right>=0$. 

2. The increment $|f(n+1)-f(n)|$ is bounded from above by some constant.

Thus, the central limit theorem gives (\ref{on}). This, together with (\ref{ineq1}) and (\ref{ineq2}) allows one to write
\be
\frac{L^{\slz}(\pi(w_n))}{n}\le \frac{L^{B_3}(w_n)}{n}\le\frac{L^{\slz}(\pi(w_n))}{n}+
O\left(\frac{1}{\sqrt{n}}\right)
\ee
Using the result $\lim_{n\to\infty}L^{\slz}(\pi(w_n))/n=1/4$ obtained 
in Appendix for symmetric random walk on $\slz$ one arrives at the following 
asymptotic expression
\be
l_{B_3}=\lim_{n\to\infty}\frac{L^{B_3}(w_n)}{n}=\frac{1}{4}
\ee

\subsubsection{Statistics of loops on $B_3$:  return probability for 
"magnetic'' random walks}

The investigation carried out above shows that if a random walk on the 
group $B_3$ ends in $Z(B_3)$ (we will say in this case a $Z$--walk), it can be regarded as a closed "magnetic" random walk on $\slz$.  Namely, if one inserts in
each elementary cell of the hyperbolic lattice a "magnetic flux" $h$ (see Fig.\ref{fig:4}) and denotes by 
$\Phi$ the total flux through a closed path on $\slz$, then any word $w^{Z}_n$ corresponding to a $Z$--walk on $B_3$  can be written as
$$
w^{Z}_n={\Delta}^{2\Phi/h}
$$ 
In other words, the group $B_3$ being the
central extension of $\slz$,  gives rise to a fibre bundle above $\slz$ such
that every full turn around the elementary cell leads to another sheet of the
Riemann surface of $\slz$.  The outcome of this construction is that $Z$--walks on $B_3$ can be decomposed into a product of elementary full turns 
around cells (this is due to the tree structure of the backbone).  
Hence the function $u_n(\Phi)$ giving the probability that a closed $n$--step 
loop on a graph $\slz$ carries a flux $\Phi$ is of great interest, especially 
because at $\Phi=0$ it defines the probability to get a {\it trivial braid} 
(i.e. completely reducible word) from a random braid of the record length $n$.

First of all we compute $u^{a}_n(\Phi)$ for a walk with local passages in a 
basis $\{a_2,b_3,a_2^{-1},b_3^{-1}\}$ (let us stress that for magnetic walks 
$a_2\not=a_{2}^{-1}$). Denote by $\sharp a_2, \sharp a_{2}^{-1}, \sharp b_3, \sharp b_{3}^{-1}$ the total 
number  of steps $a_2,a_{2}^{-1}, b_3, b_{3}^{-1}$ respectively in a given closed path on $\slz$. The fulx $\Phi$ can be written as follows:
\be\label{flux}
\Phi=\frac{h}{6}\left(3(\sharp a_2-\sharp a_{2}^{-1})+2(\sharp b_3-\sharp 
b_{3}^{-1})\right)
\ee
Recall that we consider an $n$--step process on $\slz$, conditioned by the fact 
that the path is closed (i.e. returns to the origin). Following (\ref{flux}), 
we rise a simultaneous process $\Phi_i$ (with $\Phi_0=0$) such that
\be
\Phi_{i+1}=\Phi_i+\phi_{i+1}
\ee
with $\phi_i=\pm h/2$ if the corresponding step on $\slz$ is $a_{2}^{\pm1}$, or 
$\phi_i=\pm h/3$ if the step is $b_{3}^{\pm1}$. Evidently the final 
value $\Phi_n$ gives the total flux $\Phi$ through the closed path.   

We show that the process $\Phi_i$ is not affected by the condition that the path is closed.
\begin{enumerate}
\item Notice that on $\slz$ we have $a_{2}^{-1}=a_2$, and therefore $p(h/2)=p(-h/2)$. 
\item The sign of the magnetic field  can be arbitrarily changed, hence  $p(h/3)=p(-h/3)$ (i.e. positive, $b_{3}^3$, and negative, $b_{3}^{-3}$, elementary turns are equidistributed for closed as well as for open paths).
\item The closure condition on $\slz$ affects the irreducible length of words; the irreducible forms on $\slz$ being exactly the words of the form $a^{\pm 
1}b^{\pm 1}a^{\pm 1}b^{\pm 1}a^{\pm 1}...$, setting the irreducible length of a word does not change the relative weight of $a^{\pm 1}$ and $b^{\pm 1}$ in this  word. One finally obtains
$p(h/2)=p(h/3)$. 
\end{enumerate}
The process $\Phi_i$ is then a classical one dimensional random 
walk, and therefore for $n$ large one has
\be\label{uphi}
u^{a}_n(\Phi)=\frac{1}{h\sigma_{a}\sqrt{2\pi n}}
\exp\left(-\frac{(\Phi/h)^2}{2n\sigma_{a}^2}\right)
\ee
where $\sigma_{a}^2=\frac{1}{2}\left(\frac{1}{4}+\frac{1}{9}\right)=\frac{13}{72}$.

This result seems to be interesting in the context of lattice random walks in  
a transversal magnetic field which has relations to the Harper--Hofstadter 
problem (see \cite{harper} for review) in hyperbolic geometry. 

Returning to the random walk on the braid group in the standard framing 
$\{\sigma_1, \sigma_{2}, \sigma_1^{-1}, \sigma_{2}^{-1}\}$, we can compute the 
distribution $u^{\sigma}_n(\Phi)$ for a random process on $\{{\bar 
\sigma}_1,{\bar \sigma}_2,{\bar\sigma}_1^{-1},{\bar\sigma}_2^{-1}\}$. 
Modifying slightly the derivation carried out above, one obtains that the 
corresponding process $\phi_i$ is still not affected by the condition of 
return, and is such that $p(\phi_i=h/6)=p(\phi_i=-h/6)=1/2$. This yields 
\be
u^{\sigma}_n(\Phi)=\frac{1}{h\sigma_{\sigma}\sqrt{2\pi n}}
\exp\left(-\frac{(\Phi/h)^2}{2n\sigma_{\sigma}^2}\right)
\ee
with $\sigma_{\sigma}^2=\frac{1}{36}$.

The decomposition introduced above allows to compute the return probability, 
i.e the probability $p(w_n=I_d)=p_r(n)$ to obtain a ``trivial'' braid after 
$n$ random elementary moves. Using (\ref{dec2}) the condition $w_n=I_d$ is 
equivalent to the conditions 
$$
L^{\slz}(\pi(w_n))=0\; \&\; f(n)=0
$$ 
Denote 
$$
p\{L^{\slz}(\pi(w_n))=0\}=p^{\pi}_r(n)
$$ 
and 
$$
p\{f(n)=0\ {\rm knowing}\ L^{\slz}(\pi(w_n))=0\}=p^{c}_r(n)
$$ 

The probabilities $p^{\pi}_r(n)$ and $p^{c}_r(n)$ are {\it independent} which allows 
us to set
\be \label{eq:pr1}
p_r(n)=p^{\pi}_r(n)p^{c}_r(n)
\ee
where $p^{\pi}_r(n)$ is computed in appendix A and $p^{c}_r(n)$ can be  
reexpressed the following way
\be \label{eq:pr2}
p^{c}_r(n)=hu^{\sigma}_n(0)
\ee
Collecting (\ref{eq:pr1})--(\ref{eq:pr2}) we arrive at the final expression
for $p_r(n)$
\be
p_r(n)=\frac{C}{\sigma_{\sigma}\sqrt{2\pi}}\frac{{\lam}^{n}}{n^2}
\ee
where $\lam$ and $C$ are given in appendix.

\subsubsection{Numerical results}

So far there is no constructive algorithm to find the reduced form of words of
$B_3$ for generators $\sigma_i$. The existence of an algorithm depends crucialy on the set of generators we choose. Indeed, it is shown in \cite{raz} that computing the length in terms of  generators $\sigma_i$ of a braid in $B_n$
is an NP--complete problem.  Let us mention nevertheless that braid groups are
``biautomatic'' (see \cite{char}) which basically means that there exists a set
of generators, for  which the reduced words are exactly known.  This allows in
particular to solve the word enumeration problem, and to implement methods which can compare two different braids in a polynomial time (see \cite{dehor}).
In our case of the simplest nontrivial group $B_3$ we tried a random reduction
procedure, but it converges only in exponential time.  Since our analytical
results are obtained in the regime $(n\gg 1)$, the numerical simulations give no additionnal information.

\section{Diffusion on Riemann surfaces: traces and Lyapunov exponents}
\label{sec:4}

\begin{figure}[ht]
\begin{center}
\epsfig{file=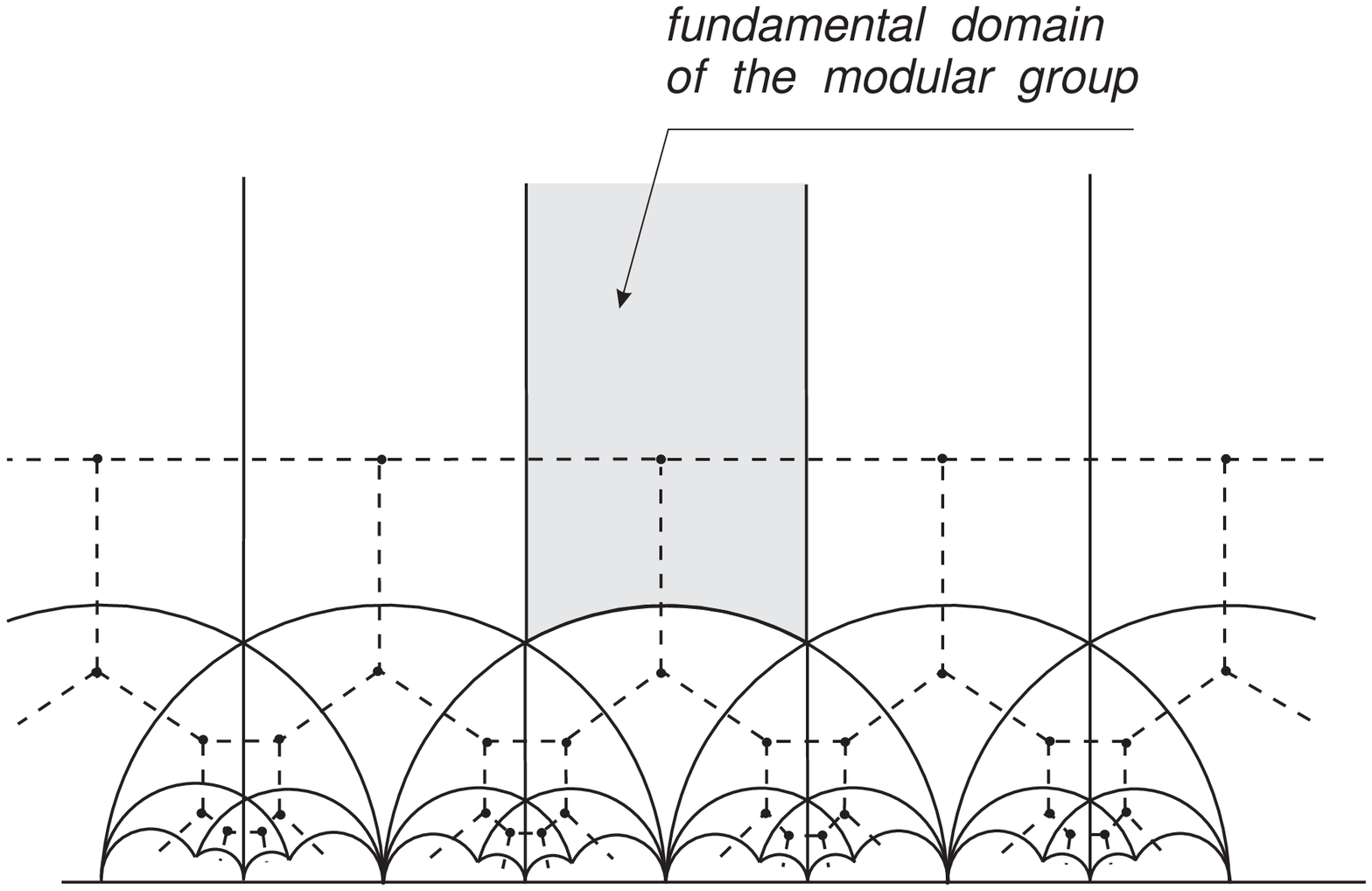,width=10cm}
\end{center}
\bigskip
\caption{Fundamental domain of  $\slz$}
\label{fig:5}
\end{figure}

We consider the representation of dimension 2 of the groups introduced above,
and investigate their action on the hyperbolic Poincar\'eé plane ${\cal H}=\{z,\ {\rm Im}\,z>0\}$.  Namely, we consider the following
fractional-linear transforms
\be
\left(\begin{array}{cc} a & b \\ c & d \end{array} \right):\
z\to\frac{az+b}{cz+d}
\ee
 We recall that $\slr$ is a subgroup of the  group of isometries of ${\cal H}$. The groups $\slz,\slz_u,H_q,F_n$ admit  representations as subgroups of $\slr$ and their Cayley graphs (considered in previous section) are now viewed as isometric lattices embedded into ${\cal H}$. Now one can investigate their metric properties using the natural hyperbolic (geodesic) distance in ${\cal H}$.  We define the lattices under consideration the same way as we have defined the Cayley graphs:
\begin{itemize}
\item We construct the set of all possible orbits of a given root point (we 
choose the point $i=(0,1)$ for conveniency) under the action of the group. 
\item We denote by $d(w_n)$ the hyperbolic distance $d(i,w_n(i))$ between $i$ 
and $w_n(i)$. 
\item We call "lattices" the Cayley graphs of the groups involved here 
because of  two important features:
\begin{itemize}
\item they are discrete subgroups of $PSL(2,\R)$, the group of motion of the 
hyperbolic 2-space.  Hyperbolic distance is a pair--point invariant, that is 
$d(i,w(i))=d(\gamma i,w\gamma(i))$, what jusities the term isometric;
\item they have the property of so-called lattice groups: they have no points 
of accumulation  (for the topology of ${\cal H}$). Recall that for $H_q$, $q\in\Z$.
\end{itemize}
\end{itemize}
Let us add that the above description  is based on well known results on Fuchsian groups theory (see \cite{katok,terras}). Properties of a Fuchsian group $G$ depend strongly on the fundamental domain of $G$, which is a minimal set of points generating ${\cal H}$ under action of $G$. The groups studied throughout this paper are all Fuchsian groups. We first  remind that the fundamental domain of the Hecke group is the circular triangle with angles $\left\{0,\frac{\pi}{q}, \frac{\pi}{q}\right\}$ (see Fig.\ref{fig:5} for $H_3$). 

\begin{figure}[ht]
\begin{center}
\epsfig{file=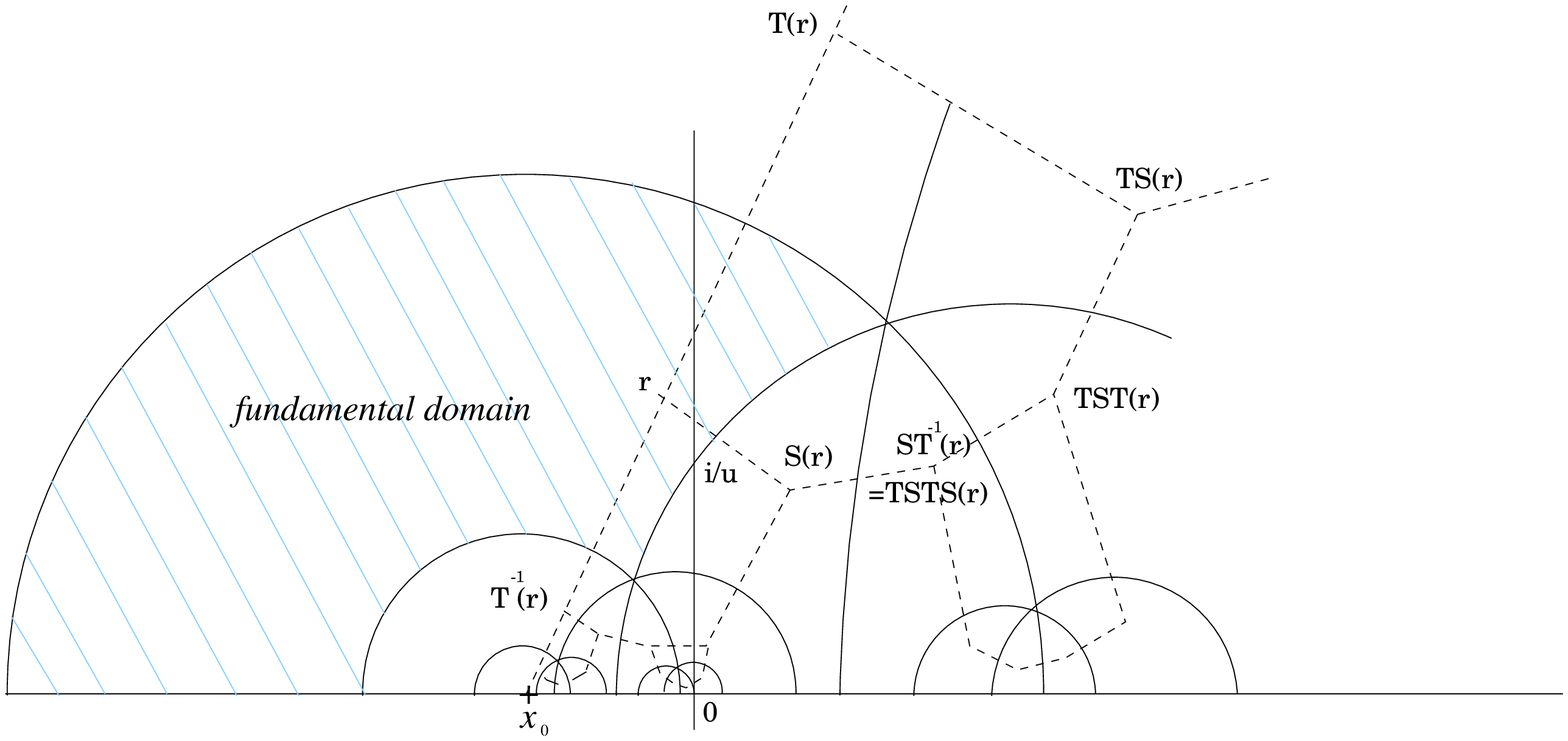,width=13cm}
\end{center}
\bigskip
\caption{Fundamental domain of  $\slz_u$, for $u=1.2$.}
\label{fig:6}
\end{figure}

It can be shown that the fundamental domain of $F_n$ is a zero-angled n-gone. Our contribution to this subject concerns the construction  of the fundamental domain of the deformed group $\slz_u$ (Fig.\ref{fig:6}). 
We omit technical details of this construction, which can be found in \cite{terras}. The outline is as follows. We first find the fixed point $i/u$ of $S_u$, and $x_0=1/(1-u^2)$ of $T_u$. We then draw the only geodesic through $i/u$ which intersects its images by $T_u$ and $T_{u}^{-1}$ with angle $\pi/3$. Circles of center $x_0$ passing by these intersections complete the construction. First notice that the topology of the Cayley graph obtained this way does not depend on $u$. Recall that only commutation relations, independent of $u$, set the topological structure of the Cayley graph). Only the metric properties are affected by $u$. In particular the area of the fundamental domain is finite only for $u=1$. The group is then said to be of type I in the classification of Fuchsians groups. For $u\not=1$ it is of type II. It means that the corresponding monodromy problems are deeply different (see \cite{bateman}). Solving the monodromy problem is an important issue since it allows to get the conformal transform that maps the fundamental domain onto ${\cal H}$. To our knowledge the problem is solved only for $u=1$. We have therefore to content ourselves with an existence theorem in the general case. Existence  of such a transform allows to define a map $f_u$ from the fundamental domain of $\slz_u$ to the fundamental domain of $\slz$. The action of $\slz_u$ on ${\cal H}$ is in this sense conjugate to the action of $\slz$:
\be
\forall\omega_u\in\slz_u,\ \omega_u(z)=f_{u}^{-1}\circ\omega_{u=1}\circ f_u(z)
\ee
The dependence on the parameter $u$ is this way clearly expressed.

\subsection{Analytic results}

Let us return to the definition of the model and recall that the groups 
under consideration act in the hyperbolic Poincar\'e upper half--plane ${\cal 
H}=\{z\in\C,\,{\rm Im}(z)>0\}$ by fractional--linear transforms\footnote{It is 
convenient first to
define the representation in the Poincar\'e upper half--plane and then use the
conformal transform to the unit disc.}. 
The matrix representation of the generators (denoted by $h_i$, $1\le i\le
n_g$) of the different groups has been given in section \ref{sec:2}.

Choosing the point $(x_0,y_0)=(0,i)$ as the tree root---see fig.\ref{fig:5},
we associate any vertex on the lattice with an element $\disp
w_n=\prod_{k=1}^n h_{\alpha_k}$ where $1\le\alpha_k \le n_g$ and $w_n$ is
parametrized by its complex coordinates $z_n=w_{n}\big(i\big)$ in the
hyperbolic plane.

Strictly speaking ${\cal H}$ should be identified with $SL(2,\R)/SO(2)$; we here identify an element with its class of equivalence of $SO(2)$. The following
identity holds (see \cite{terras,els})
\be \label{eq:cosh} 
2\cosh\Big(d(w_n)\Big)={\rm Tr}(w_{n}w_{n}^{\dag}) 
\ee 
where dagger denotes transposition.

We are interested in the distribution function $P_n(d)$, and therefore have to 
look for the distribution of traces of matrices $w_n$. The method described 
hereafter involves mainly the results of the paper \cite{CNV}. The outline of our 
approach is as follows. We study the behavior of the random matrix $w_n$, generated by a Markov chain (which must fulfill ergodicity properties) defined as follows: 
\be \label{mar}
w_{n+1}=M_n h_{\alpha_{n+1}}{\rm ,}\ (\alpha_{n+1}=i,\ 1\le i\le n_g)\ 
{\rm with\  probability}\ \frac{1}{n_g}
\ee

We use the standard methods of random matrices and consider the entries of the
$2\times  2$--matrix $w_n$ as a 4--vector ${\cal V}_n$. The transformation
$w_{n+1}= w_n h_{\alpha}$ reads now
\be
{\cal V}_{n+1}= \left(\begin{array}{ll} h_{\alpha}^{\dag} & 0 \\ 0 &
h_{\alpha}^{\dag} \end{array}\right)\; {\cal V}_{n}
\ee
This block--diagonal form allows to  study  one
of two 2--vectors composing ${\cal V}_{n}$, say $v_n$. Parametrizing
$v_n=(\varrho_n\cos\theta_n,\varrho_n\sin\theta_n)$ and using the relation
$d(w_n)\equiv  d_n\simeq 2\ln\varrho_n$ valid for $n\gg 1$, one gets a
recursion relation $v_{n+1}=h_{\alpha}^{\dag}v_n$ in terms of hyperbolic
distance $d_n$:
\be \label{eq:dist}
\label{dist}
\disp  d_{n+1}= d_n+\ln\,
p_{\al}(\cos\theta)
\ee
where $p_{\alpha}$ is a second order polynomial depending on the specific form of transition matrices $h_{\alpha}$. While for the angles one gets straightforwardly
\be \label{ang}
\cot\theta_{n+1}=h_{\alpha}\left(\cot\theta_n\right)
\ee
The action of $h_{\alpha}$ is fractional--linear.

One has now to study the invariant measure $\mu(\theta)$, giving the asymptotic
probability to have $\theta_n=\theta$.
Introducing $x=\cot\theta$, we are led to study the action of the group
restricted on the real line parametrized by $x$. The statistical properties of
$\mu$ have been discussed by Gutzwiller and Mandelbrot \cite{gut} in the case 
of the free group $\Lambda$.  An alternative, put forward in \cite{CNV}, is to
define $\mu(x)$ as the limit of the following recursion relation:
\be \label{rec2}
\mu^{(n+1)}(x)=\frac{1}{n_g}
\sum_{\alpha=1}^{n_g}\mu^{(n)}\Big(h_{\alpha}(x)\Big)
\left|\frac{dh_{\alpha}(x)}{dx}\right|
\ee


The convergence $\mu^{(n)}(x)\to\mu(x)$ for $n\to\infty$ is assured
by ergodic properties of the functional transform (\ref{rec2}) and has been successfully checked numerically by comparing to direct sampling of different
orbits. Despite the absence of  rigorous proof, we claim that $\mu$ is  defined with no ambiguity by (\ref{rec2}). This enables us to compute the desired
distribution $P_n(d)$. The crucial point required for convergence of
$\mu^{(n)}$ to the invariant distribution, is the existence of ergodic properties of
$\theta_n$. It means that for $n\gg 1$, the distribution of $\theta_n$ is
exactly given by $\mu(\theta)$, independently  of $n$ and initial conditions. We  introduce the generating function for (\ref{dist}); due to the Markovian structure of (\ref{dist}), we can perform the averaging:
\be
\left<e^{ikd_{n+1}}\right>=
\left<e^{ikd_n}\right>\left<[p_{\al}(\cos\theta)]^{ik}\right>
\ee
Thus we obtain
\be
\left<e^{ikd_n}\right>=
\left[\frac{1}{n_g}\sum_{\al=1}^{n_g}\int_{-\pi/2}^{\pi/2}
d\theta\mu(\theta)[p_{\al}(\cos\theta)]^{ik}\right]^n
\ee

This form suggests that for $n$ large $P_n(d)$ satisfies a central limit
theorem. Indeed such a theorem exists (see \cite{doob,boug}) for 
Markovian processes provided that the phase space is ergodic. We are then 
led to compute only the first two moments (Lyapunov exponents), which gives
us
\be
\gamma_1=\lim_{n\to\infty}\frac{\left<d\right>}{n}=
\frac{1}{n_g}\sum_{\al=1}^{n_g}\int_{-\pi/2}^{\pi/2}
d\theta\mu(\theta)\ln p_{\al}(\cos\theta)
\ee
and
\be
\sigma^2=\lim_{n\to\infty}\frac{\left<(d-\left<d\right>)^2\right>}{n}=
\gamma_2-\gamma_1^2
\ee
where
\be
\gamma_2=\frac{1}{n_g}\sum_{\al=1}^{n_g}\int_{-\pi/2}^{\pi/2}
d\theta\mu(\theta)\ln^2 p_{\al}(\cos\theta)
\ee

\subsection{Numerical results}

We present in this part the numerical and semi-analytical results for the 
invariant measure $\mu$ and the Lyapounov exponent $\gamma_1$. Our main goal is
to compare the approach developed here with  the results following from  the study of random walks on graphs (see section \ref{sec:3}.  

Let us call the {\it backbone subgroup} ${\cal B}(G)$ of the group $G$ such
subgroup of $G$ whose Cayley graph is the backbone of the graph of $G$. It seems to be more instructive to rely on this purely geometrical caracterization of ${\cal B}$ and to avoid a formal formal definition. Let us stress 
that ${\cal B}(G)$ is a free subgroup of $G$. One has for example ${\cal 
B}(F_n)=F_n$. Consider now the representation of $F_q$ by $q$ 
idempotent generators $g_1,..,g_q$ with the following homomorphism $\Psi$:
\be
\Psi:\left\{\begin{array}{lll}
F_q & \longrightarrow & H_q\\
g_i &\longrightarrow & b_{q}^{-i}a_2b_{q}^{i}
\end{array}\right.
\ee
Due to injectivity of $\Psi$, the following decomposition holds
\be
H_q=\bigcup_{i=1}^{q}b_{q}^{i}\Psi(F_q)
\ee
with
\be
b_{q}^{i}\Psi(F_q)\bigcap b_{q}^{j}\Psi(F_q)=\emptyset\ {\rm for}\ i\not=j
\ee
what means that the Cayley graph of $H_q$ is the disjoint union of $q$ trees
$\T_q$. Thus we set ${\cal B}(H_q)=F_q$.

The scale factor $s_f$ is the ``average'' irreducible length of the generators 
of ${\cal B}(G)$ in $G$. In other words, $L^G(w)\sim s_f L^{{\cal B}(G)}(w)$ 
for $w\in{\cal B}(G)$ with $L^G(w)\gg 1$. We have studied two different
Markovian processes for each group $G$: (i) simple random walks 
(characterized by the Lyapunov exponent $\gamma_{1}^{s}$) and (ii) so-called {\it directed} 
random walks (that are walks excluding two consecutive opposite steps) on the
backbone subgroup ${\cal B}(G)$ (caracterized by Lyapunov exponent $\gamma_{1}^{d}$). 

By construction $\gamma_{1}^{d}/s_f$ gives the average hyperbolic length 
for an elementary step on $G$. We conjecture that $s_{f}\gamma_{1}^{s}/
\gamma_{1}^{d}$ gives {\it the number of steps to the origin} (normalized 
by $n$) on the graph $G$. Let us point out that this result links together 
two definitions of the "drift" for  random walks on the groups $G$: the drift $l$ is defined on the graph in metric of words while $\gamma$ is defined in terms of hyperbolic distance for an isometric embedding of $G$ into ${\cal
 H}$. Thus we claim
\be \label{eq:relat}
\left<L^{G}(w)\right>=\frac{s_f}{\gamma_{1}^{d}}\left<\ln{\rm 
Tr}(ww^{\dag})\right>
\ee
where a word $w$ is identified with its matrix representation. We believe that equation \ref{eq:relat} is worth interest, since it relates properties of a group defined only trough symbolic commutation relations, to geometrical properties of a given representation. 

The stochastic average $\left<...\right>$ in (\ref{eq:relat}) is necessary, to
wash out purely geometrical effects such as multifractality investigated in
\cite{CNV}.  (It corresponds to fluctuations of the hyperbolic distance for words
of same length on the backbone graph).  One  has to stress  that (\ref{eq:relat}) holds due to a ``global'' spherical symmetry (see \cite{lyons} for a precise definition of this symmetry for graphs) of both models;
only the ``radial'' part of the processes is considered, whereas the angular
dependence is averaged (here again the ergodic properties play the crucial
role).  This has been checked numerically in the continuous case:  generators
have to be properly normalized, such that each elementary step should have the
same hyperbolic length, ensuring spherical symmetry, else the invariant measure
$\mu$ fails to converge.

All results are summarized in table \ref{tab:1}.

\begin{table}
\begin{center}
\begin{tabular}{|c|c|c|c|c|c|}
\hline 
group  & generators & ${\mbox{backbone subgroup}, \atop \mbox{scale factor
$s_f$}}$ & ${s_{f}\gamma_{1}^{s}/\gamma_{1}^{d} \atop \mbox{numerical}}$ &
${s_f\gamma_{1}^{s}/\gamma_{1}^{d} \atop \mbox{semi--analytical}}$ & ${\left<d\right>/n: \atop \mbox{graph approach}}$ \\ \hline \hline
$F_3$ & $h_1,h_2,h_3$ & $F_3,\ 1$ & 0.3334 & 0.332 & 1/3 \\ \hline
$F_4$ & $h_1,h_2,h_{1}^{-1},h_{2}^{-1}$ & $F_4,\ 1$ & 0.501 & 0.503 & 1/2 \\ 
\hline $H_3$ & $a_2,b_3,b_{3}^{-1}$ & $F_3$,\ 2 & 0.1334 & 0.132 &
$2/15=0.133..$ \\ \hline $\slz$ & $\bar{\sigma}_1,\bar{\sigma}_2,
\bar{\sigma}_{1}^{-1}, \bar{\sigma}_{2}^{-1}$ & $F_3$,\ 1 & 0.2501 & 0.248 &
$1/4$ \\ \hline \hline 
\end{tabular}
\end{center}
\caption{}
\label{tab:1}
\end{table}

\section{Discussion and perspectives}
\label{sec:5}
We have presented during this work different aspects of random walks on a
family of hyperbolic groups.  On one hand we studied the Cayley graphs of these
groups, and briefly exposed general methods of computing   the Green
functions for Markovian processes on those graphs; in particular we explicitly calculate the drift in different cases.  As an application, we studied Markovian processes on the braid group $B_3$, and explicitly showed that the
drift for a symmetric random walk on this group tends at $n\to\infty$ to the drift of a process on the group $\slz$, which is found to be $1/4$. This means that a typical random braid of record length $n$ can be released on average by  $n/4$ elementary moves. The graph approach and the introduction of ``magnetic walks''  enabled us also to compute explicitely the return
probability on $B_3$, that is the probability to obtain a trivial (completely reducible) braid from a random word  of record length $n$.

On the other hand we took advantage of the fact that the groups $H_q$ and $\slz_u$ are subgroups of $\slr$ and therefore act naturally in the hyperbolic plane $\cal
H$.  The Cayley graphs of these groups are then naturally embedded in $\cal H$.
Instead of the usual length in metric of word, we could, thanks to this representation, use
the metric structure of $\cal H$ and study the hyperbolic length of random
elements of the group.  This problem leads to the  study of products of random
matrices.  The method described in \cite{CNV}  allows us to compute the
probability distribution of the hyperbolic length.  Lyapunov exponents are
explicitely computed in different cases.

These two approaches are shown to be related by an equation (\ref{eq:relat}).
This result is a strong motivation for investigating further the geometric
properties of hyperbolic groups in connexion with other topological invariants.
As an example we briefly mention the Alexander polynomials.

The Alexander polynomial $\nabla_K(t)$ of a link $K$ represented by a closed
braid $w_n=\prod_{j=1}^{n}\sigma_{r_j}$ of length $n$ is defined as follows
\be \label{alex}
(1+t+t^2)\,\nabla_K(t) = \det\left[\prod_{j=1}^{n} 
\hat{\sigma}_{r_j}-\hat{I}\right] = 
\det\left[\prod_{j=1}^{n}\hat{\sigma}_{r_j}\right] +1 -
{\rm Tr}\left[\prod_{j=1}^{n}\hat{\sigma}_{r_j}\right]
\ee
where $j$ runs ``along the braid", i.e. labels the number of used
generators, the subscript $r_{j}\in\{-2,-1,1,2\}$ marks the set of braid 
generators (letters), with the prescription 
$\hat{\sigma_i}^{-1}=\hat{\sigma}_{-i}$ and
$\hat{I}$ defines the $2\times 2$--identity matrix. For long words ($n\gg1$), the 
following asymptotic expression  holds:
\be
{\rm Tr}(w_n)\sim\left({\rm Tr}(w_nw_{n}^{\dag})\right)^{1/2}\sim e^{d(w_n)/2}
\ee
One then has, with the parameter $u=\sqrt{-t}$ (recall that $\hat{\sigma_i}$ 
depends on $u$):
\be \label{alex2}
(1-u^2+u^4)\,\nabla_K(u)=u^{2p(w_n)}-u^{p(w_n)}e^{d(w_n)/2}+1
\ee
with $p(w_n)=\sharp(+)-\sharp(-)$. In this regime the polynomial is therefore 
expressed only in terms of $p(w_n)$ and $d(w_n)$. The quantity $p(w_n)$ is a ``poor'' 
invariant, in sense that it takes the same value for a large amount of links. In 
other words  $p(w_n)$ is just the length of the element $w_n$ projected onto $\Z$. 
Indeed there exists an obvious group homomorphism $\pi_1$ from $B_3$ to $\Z$
defined by $\pi_1(\sigma_{i}^{\pm1})=\pm1$.  All non abelian properties are
lost by this invariant.  The geometric invariant $d(w_n)$, described above, is
much stronger.  As we have shown, this invariant is related directly to the
word length in the group $\slz$, which preserves the noncommutative structure
of $B_3$ (recall that the random word length in $B_3$ has the same asymptotics
as the word length in $\slz$).  The information is nevertheless not
redundant, because there is no nontrivial homorphism from $\slz$ to $\Z$
(there is no finite order element in $\Z$).  In particular, under the condition
$d(w_n)=0$ (Z--walks), $p$ is an exact invariant having sense of a winding number.

The form (\ref{alex2}) seems in particular convenient for possible problems of
statistics of Alexander polynomials, since we know the statistics of both $p$
and $d$.  In particular, for a simple random walk, the typical Alexander
polynomial could be defined as ${\bar\nabla}_n(u)$:
\be
(1-u^2+u^4){\bar\nabla}_n(u)=1-e^{n\gamma_1(u)/2}
\ee
where $\gamma_1(u)$ is the Lyapunov exponent of the random product of generators 
$\hat\sigma_i$.

\begin{appendix}
\section{Drift on $\slz$}
\label{app1}
\begin{figure}[ht]
\begin{center}
\epsfig{file=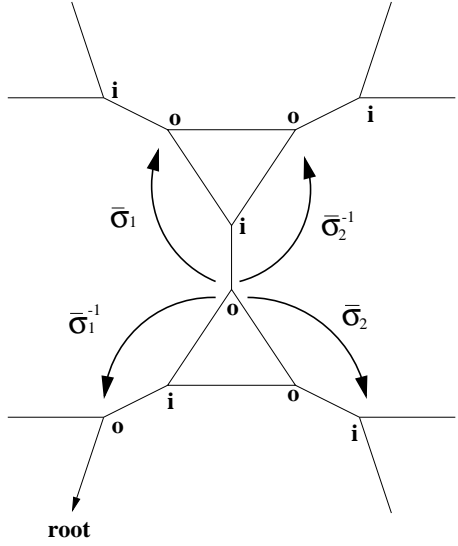,width=6cm}
\end{center}
\bigskip

\caption{Random walk on $\slz$
in terms of generators $\bar \sigma_i$. Vertices of type i and o are shown.}
\label{fig:7}
\end{figure}

The goal of this appendix is to compute the drift of  a random walk on $\slz$
in terms of generators $\bar \sigma_i$. We keep notations of \ref{sec:3} and 
proceed the same way, noting that the process under consideration is no longer a 
simple random walk, but is described by the transitions shown in fig.\ref{fig:7}.

A direct counting gives the following master equation for $k\ge2$:
\be\label{maeq1}
\left\{\begin{array}{l}
{\bar P}_{3}^{i}(k,n+1)=\frac{1}{4}\left({\bar P}_{3}^{i}(k+1,n)+2{\bar
P}_{3}^{i}(k-1,n)+{\bar P}_{3}^{o}(k-1,n)\right) \medskip \\ 
{\bar P}_{3}^{o}(k,n+1)=\frac{1}{4}\left({\bar P}_{3}^{o}(k+1,n)+{\bar
P}_{3}^{i}(k+1,n)+2{\bar P}_{3}^{o}(k-1,n)\right)
\end{array}\right.
\ee
with initial conditions of the form
\be\left\{\begin{array}{l}
{\bar P}_{3}^{i}(k,0)=\alpha\delta_{k,0} \medskip \\
{\bar P}_{3}^{o}(k,0)=(1-\alpha)\delta_{k,0}
\end{array}\right.
\end{equation}

One then straightforwardly obtains the following algebraic linear system:
\be\label{alsys1}
\left\{\begin{array}{l}
\disp Q^{i}(x,s)\left(1-\frac{s}{4}(e^{-ix}+2e^{ix})\right)-
\frac{s}{4}e^{ix}Q^{o}(x,s)= \alpha \medskip \\
\disp -\frac{s}{4}e^{-ix}Q^{i}(x,s)+
\left(1-\frac{s}{4}(e^{-ix}+2e^{ix})\right)Q^{o}(x,s)=1-\alpha
\end{array}\right.
\ee
determining $Q^{i,o}(x,s)$:
\be
\disp Q^{i,o}(x,s)=\frac{a_{\alpha}^{i,o}(x)+b_{\alpha}^{i,o}(x)s}{p(x,s)}
\ee

We omit the details irrelevant for the purpose of this Appendix.
We denote as $s_{\pm}(x)$ the roots of $p(x,s)$. They obey the equations
\be
\disp\left\{\begin{array}{l}
\disp s_{+}=2-ix+O(x^2) \medskip \\
\disp s_{-}=1-\frac{ix}{4}+O(x^2)
\end{array}\right.
\end{equation}
and one gets finally
\be 
\frac{\langle k\rangle_3}{n}=\frac{1}{4}
\ee

\begin{figure}[ht]
\begin{center}
\epsfig{file=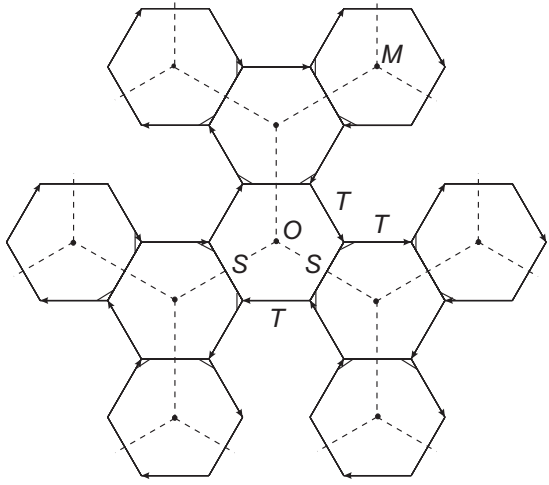,width=8cm}
\end{center}

\caption{The honey-comb lattice.}
\label{fig:8}
\end{figure}

One has now to make sure that for any word $w_n$ of $n$ letters in the alphabet
${\bar \sigma_i}$ the following relation holds:
\be \label{a1}
k(w_n)=L(w_n)+O(1)
\ee
Even if Fig.\ref{fig:7} makes this statement clear, a more rigorous proof is
as follows. Consider a given word $w$, with $k(w)=k_0$. Then the following 
decomposition holds:
\be
w=b_{3}^{\epsilon_0}\left(\prod_{i=1}^{k_0}a_2b_{3}^{\epsilon_i}\right)
a_{2}^{\epsilon_f}
\ee
with $\epsilon_0\in\{0,1,2\},\ \epsilon_i\in\{1,2\},\  \epsilon_f\in\{0,1\}$.
To prove (\ref{a1}) we use the relation
\be
L\left(\prod_{i=1}^{k_0}a_2b_{3}^{\epsilon_i}\right)=k_0
\ee
and one has finally
\be
\lim_{n\to\infty}\frac{L(w_n)}{n}=\frac{1}{4}
\ee
Let us mention that a more direct derivation of this result can be brought in
if one considers the $\slz$ generators . The structure of the Cayley graph of
$\slz$ depends on the basis and in the framing $S,T,T^{-1}$ it has form
of the so-called hyperbolic honeycomb lattice (see Fig.\ref{fig:8}). 

Define $\kappa$---the distance on the backbone graph of $\slz$.  The partition function $P_n(\kappa)$ 
to find the random walker at a distance in $\kappa$ steps along the backbone graph from the origin after $n$ elementary steps satisfies the master equation:
\be
P_{n+1}(\kappa)=\frac{1}{4}P_n(\kappa+1)+\frac{1}{4}P_n(\kappa)+
\frac{1}{2}P_n(\kappa-1)
\ee
with the following  boundary conditions:
\be
P_{n+1}(0)=\frac{1}{2}(P_{n}(0)+P_{n}(1))
\ee
This is a standard problem whose solution is known, and the condition $L(w_n)=0$ is in 
particular equivalent to $\kappa=0$, therefore the probability to obtain a trivial word after $n$ random steps (denoted $p^{\pi}_r(n)$) is given by
\be
p^{\pi}_r(n)=P_n(0)=C\frac{{\lam}^n}{n^{3/2}}
\ee
with 
$$
C=\frac{9+4\sqrt{2}}{7\pi}\ \mbox{and}\ \lam=\frac{2\sqrt{2}+1}{4}
$$

\end{appendix}

\end{document}